\documentclass[conference]{IEEEtran}
\usepackage{epsfig}
\usepackage{times}
\usepackage{float}
\usepackage{afterpage}
\usepackage{amsmath}
\usepackage{amstext}
\usepackage{amssymb,bm}
\usepackage{latexsym}
\usepackage{color}
\usepackage{graphicx}
\usepackage{amsmath}
\usepackage{amsthm}
\usepackage{graphicx}
\usepackage[center]{caption}
\usepackage{pstricks}
\usepackage{subfigure}
\usepackage{booktabs}
\usepackage{multicol}
\usepackage{lipsum}
\usepackage{dblfloatfix}
\allowdisplaybreaks

\newtheorem{thm}{Theorem}

\newtheorem{prop}[thm]{Proposition}

\newtheorem{defin}{Definition}

\begin{document}

\title{The Approximate Optimality of Simple Schedules for Half-Duplex Multi-Relay Networks}

\author{
\IEEEauthorblockN{Martina Cardone$^{\dagger}$, Daniela Tuninetti$^*$ and Raymond Knopp$^{\dagger}$}
$^{\dagger}$Eurecom,
Biot, 06410, France, 
Email: \{cardone, knopp\}@eurecom.fr\\
$^*$ University of Illinois at Chicago,
Chicago, IL 60607, USA, 
Email: danielat@uic.edu}\maketitle
\begin{abstract}
In ISIT'12 Brahma, \"{O}zg\"{u}r and Fragouli 
conjectured that in a half-duplex diamond relay network (a Gaussian noise network without a direct source-destination link and with $N$ non-interfering relays) an approximately optimal relay scheduling (achieving the cut-set upper bound to within a constant gap uniformly over all channel gains) exists with at most $N+1$ {\it active} states (only $N+1$ out of the $2^N$ possible relay listen-transmit configurations have a strictly positive probability). Such relay scheduling policies are said to be {\it simple}. 
In ITW'13 we conjectured that simple relay  policies are optimal for any half-duplex Gaussian multi-relay network, that is, simple schedules are not a consequence of the diamond network's sparse topology. 
In this paper we formally prove the conjecture beyond Gaussian networks. In particular, for any memoryless half-duplex $N$-relay network with independent noises and for which independent inputs are approximately optimal in the cut-set upper bound, an optimal schedule exists with at most $N+1$ active states. 
The key step of our proof is to write the minimum of a submodular function by means of its Lov\'{a}sz extension and use the greedy algorithm for submodular polyhedra to highlight structural properties of the optimal solution. This, together with the saddle-point property of min-max problems and the existence of optimal basic feasible solutions in linear programs, proves the claim.
\end{abstract}

\section{Introduction}
Adding relaying stations to today's cellular infrastructure promises to boost both coverage and network throughput.
Although higher performances could be attained with Full-Duplex (FD) relays, due to practical restrictions, such as the inability to perfectly cancel the self-interference, currently employed relays operate in Half-Duplex (HD).

This paper studies a general memoryless multi-relay network, where the communication between a source and a destination is assisted by $N$ relays operating in HD mode. 
The capacity of this network is not known in general.
In~\cite{ourITjournal} we 
showed that Noisy Network Coding (NNC)~\cite{nncLim} achieves the cut-set upper bound~\cite{book:ElGamalKim2012} to within $1.96\left(N+2\right)$ bits per channel use for a general Gaussian noise multi-relay network, universally over all channel gains, 
thus improving on previously known constant gap results.
In general, finding the capacity of a HD multi-relay network is a combinatorial problem since the cut-set upper bound is the minimum between $2^N$ bounds (one for each possible cut in the network), each of which is a linear combination of $2^N$ relay states (since each relay can either transmit or receive). Thus, as the number of relays increases, optimizing the cut-set bound becomes prohibitively complex. Identifying structural properties of the cut-set upper bound, or of a  constant gap approximation of the cut-set upper bound, is therefore critical for efficient numerical evaluations and can have important practical consequences for the design of reduced complexity / simple relaying policies.

In~\cite{Bagheri2009}, the authors analyzed the Gaussian HD diamond relay network, 
a multi-relay network without a direct source-destination link, with $N=2$ non-interfering relays
and proved that at most $N+1=3$ states, out of the $2^N=4$ possible ones, suffice to 
characterize the capacity to within a constant gap.
We say that 
these $N+1$ states are {\it active} and form an (approximately optimal) {\it simple} schedule.
In~\cite{Fragouli2012}, Brahma {\it et al} verified through extensive numerical evaluations that in Gaussian HD diamond networks with $N \leq 7$ relays an optimal (to within a constant gap) schedule has at most $N+1$ active states and
conjectured this to be true for any $N$. 
In~\cite{BrahmaISIT2014}, Brahma {\it et al}'s conjecture was proved for Gaussian HD diamond networks with $N \leq 6$ relays; the proof is based on certain properties of submodularity and on linear programming duality; the proof technique does not appear to easily generalize to an arbitrary $N$.
%
Our numerical experiments in~\cite{ourITW2014} showed that Brahma {\it et al}'s conjecture on the existence of optimal simple schedules for diamond HD relay networks extends to any Gaussian HD multi-relay network (i.e., not necessarily with a diamond topology) with $N \leq 8$;
we conjectured that the same holds for any $N$. Should our more general version of  Brahma {\it et al}'s conjecture be true, then Gaussian HD multi-relay networks have optimal simple schedules irrespectively of their topology.
In~\cite{ourITjournal} we discussed polynomial time algorithms to determine the optimal simple schedule and extensions beyond relay networks.

Related works on determining the optimal relay scheduling, but not focused on characterizing the minimum number of active states, are available in the literature. 
For example~\cite{OngMultiRelay} studied an iterative algorithm to determine the optimal schedule when the relays use decode-and-forward. 
In~\cite{EtkinParvareshShomoronyAvestimehr} the authors proposed a `grouping' technique to compute the relay schedule that maximizes the approximate capacity of certain Gaussian HD relay networks; 
because finding a good node grouping is computationally complex, the authors proposed a heuristic approach based on tree decomposition which results in polynomial time algorithms; as for diamond networks in~\cite{Fragouli2012}, 
the low-complexity algorithm of~\cite{EtkinParvareshShomoronyAvestimehr} relies on the `simplified' topology of certain networks.
{\it As opposed to these works, we prove that a linear number of states is sufficient to determine an optimal schedule regardless of the network topology.
}
We also note that in~\cite{ParvaEtk}, FD relay networks were studied and that ``under the assumption of independent inputs and noises, the cut-set bound is submodular''~\cite[ Theorem 1]{ParvaEtk}, a result that we shall use in the derivation of our main result.

The main result of this paper is a formal proof of Brahma {\it et al}'s conjecture beyond the Gaussian noise case.
In particular, we prove that for any HD network with $N$ relays, with independent noises and for which independent inputs in the cut-set bound are approximately optimal, the optimal relay policy is simple. 
The key idea is to use the Lov\'{a}sz extension and the greedy algorithm for submodular polyhedra to highlight structural properties of the minimum of a submodular function. Then, by using the saddle-point property of min-max problems and the existence of optimal basic feasible solutions for Linear Programs (LPs), an (approximately) optimal relay policy with the claimed number of active states can be shown. A polynomial time algorithm to find the optimal simple relay schedule is also discussed.

The rest of the paper is organized as follows. 
Section \ref{sec:sysModel} describes the general memoryless HD multi-relay network. 
Section \ref{sec:mainRes} summarizes some known results for submodular functions and LPs and then proves the main result.
Finally, Section \ref{sec:Concl} concludes the paper.

\section{System Model}
\label{sec:sysModel}

A memoryless relay network has one source (node 0), one destination (node $N+1$), and $N$ relays (indexed from $1$ to $N$). It consists of $ N+1$ input alphabets $\left (\mathcal{X}_1,\cdots,\mathcal{X}_{N},\mathcal{X}_{N+1} \right )$ (here $\mathcal{X}_i$ is the input alphabet of node~$i$ except for the source / node~0 where, for notation convenience, we use $\mathcal{X}_{N+1}$ rather than $\mathcal{X}_{0}$), $N+1$ output alphabets $\left (\mathcal{Y}_{1},\cdots,\mathcal{Y}_{N},\mathcal{Y}_{N+1} \right )$ (here $\mathcal{Y}_i$ is the output alphabet of node~$i$), and a memoryless channel transition probability $\mathbb{P}_{Y_{[1:N+1]}|X_{[1:N+1]}}$. 
Codes, achievable rates and capacity are defined in the usual way (see for example~\cite{ourITjournal}).

In this general memoryless framework, each relay can listen and transmit at the same time, i.e., it is a FD node.
HD channels are a special case of the memoryless FD framework in the following sense~\cite{kramer-allerton}.
With a slight abuse of notation compared to the previous paragraph, we let the channel input of the $k$-th relay, $k\in[1: N]$, be the pair $(X_k,S_k)$, where $X_k\in \mathcal{X}_k$ as before and $S_k \in  [0:1]$ is the {\em state} random variable that indicates whether the $k$-th relay is in receive-mode ($S_k=0$) or in transmit-mode ($S_k=1$).
In the HD case the channel transition probability is specified as $\mathbb{P}_{Y_{[1:N+1]}|X_{[1:N+1]},S_{[1:N]}}$. In particular, when the $k$-th relay, $k\in [1:N]$, is listening ($S_k=0$) the outputs are independent of $X_k$, while when the $k$-th relay is transmitting ($S_k=1$) its output $Y_k$ is independent of all other random variables. 

The capacity $\mathsf{C}$ of the HD multi-relay network is not known in general, but can be upper bounded by the cut-set bound 
\begin{align}
\mathsf{C} & \leq 
{ \max_{\mathbb{P}_{X_{[1:N+1]},S_{[1:N]}}}} \min_{\mathcal{A} \subseteq [1:N]} I_{\mathcal{A}}^{(\text{rand})},
\label{eq:capcutsetup}
\quad \text{where}
\\ 
I_{\mathcal{A}}^{(\text{rand})} &:=
I \left( X_{N + 1}, X_{\mathcal{A}^c},{ S_{\mathcal{A}^c}};
Y_{N + 1},Y_{\mathcal{A}} |X_{\mathcal{A}}, 
{ S_{\mathcal{A}}}
\right)
\label{eq:Iranddef}
\\&
\leq H({ S_{\mathcal{A}^c}}) + I_{\mathcal{A}}^{(\text{fix})}, 
\label{eq:Iranddefup}
\\
I_{\mathcal{A}}^{(\text{fix})} &:=
I \left( X_{N + 1}, X_{\mathcal{A}^c};
Y_{N + 1},Y_{\mathcal{A}} |X_{\mathcal{A}},
{ S_{[1:N]}}
\right)
\label{eq:IAfixed}
\\&= \sum_{s  \in  [0:1]^N} \lambda_s \ f_s(\mathcal{A}),
\ \text{for} \ \lambda_{s}:= \mathbb{P}[S_{[1: N]}=s] \ \text{and}
\\
f_s(\mathcal{A}) &:=
I \left( X_{N + 1}, X_{\mathcal{A}^c};
Y_{N + 1},Y_{\mathcal{A}} |X_{\mathcal{A}},  
{ S_{[1:N]} =s }
\right).
\label{eq:fs}
\end{align}
In the following, we use interchangeably the notation  $s  \in  [0:1]^N$ to index all possible binary vectors of length $N$, as well as, $s\in [0:2^N-1]$ to indicate the decimal representation of a binary vector of length $N$.
$I_{\mathcal{A}}^{(\text{rand})}$ in~\eqref{eq:Iranddef} is the mutual information across the network cut $\mathcal{A}\subseteq[1:N]$ when a {\it random schedule} is employed, i.e., information is conveyed from the relays to the destination by switching between listen and transmit modes of operation at random times~\cite{kramer-allerton} (see the term  $H({ S_{\mathcal{A}^c}})\leq \left |\mathcal{A}^c \right| \leq N$ in~\eqref{eq:Iranddefup}).
$I_{\mathcal{A}}^{(\text{fix})}$ in~\eqref{eq:IAfixed} is the mutual information with a {\it fixed schedule}, i.e., the time instants at which a relay transitions between listen and transmit modes of operation are fixed and known to all nodes in the network~\cite{kramer-allerton} (see the term $S_{[1:N]}$ in the conditioning in~\eqref{eq:IAfixed}).
Note that fixed schedules are optimal to within $N$ bits.

\section{Main Result}
\label{sec:mainRes}

\begin{subequations}
We next consider networks for which the following holds:
there exists a product input distribution
\begin{align}
\mathbb{P}_{X_{[1:N+1]}|S_{[1:N]}} 
= \prod_{i\in[1:N+1]} \mathbb{P}_{X_i|S_{[1:N]}}
\label{eq:indipinputs}
\end{align}
for which we can evaluate the set function $I_{\mathcal{A}}^{(\text{fix})}$ in~\eqref{eq:IAfixed}
for all $\mathcal{A}\subseteq[1:N]$ and bound the capacity as
\begin{align}
\mathsf{C}^\prime-\mathsf{G}_1 &\leq  \mathsf{C}  \leq \mathsf{C}^\prime+\mathsf{G}_2, 
: \ \mathsf{C}^\prime := \max_{\mathbb{P}_{S_{[1:N]}} }  
\min_{\mathcal{A} \subseteq [1:N]}
I_{\mathcal{A}}^{(\text{fix})},
\label{eq:capinidinputs}
\end{align}
and where $\mathsf{G}_1$ and $\mathsf{G}_2$ are non-negative constants that may depend on $N$ but not on the channel transition probability.
In other words, we concentrate on networks for which using independent inputs and a fixed relay schedule in the cut-set bound provides both an upper bound, to within $\mathsf{G}_2$ bits, and a lower bound, to within $\mathsf{G}_1$ bits, on the capacity.

For example, for a general Gaussian multi-relay network with independent noises,  independent Gaussian inputs are optimal to within $\mathsf{G}_1 +\mathsf{G}_2 \leq  1.96 (N+2)$~bits universally over all channel gains~\cite{ourITjournal}.
In \cite{ourITjournal} we conjectured that the optimal schedule in this case would be a simple one, i.e.,  the optimal probability mass function $\mathbb{P}_{S_{[1:N]}}$ in~\eqref{eq:capinidinputs} is such that at most $N+1$ entries have a strictly positive probability. This paper proves that  not only the conjecture is true for the Gaussian noise case, but it also holds in more generality.

The main result of the paper is:
%
\begin{thm}
Under the assumptions in~\eqref{eq:allassumptions} and of 
\begin{align}
&\mathbb{P}_{Y_{[1:N+1]}|X_{[1:N+1]},S_{[1:N]}} = 
\prod_{i\in[1:N+1]} \mathbb{P}_{Y_{i}|X_{[1:N+1]},S_{[1:N]}},
\label{eq:indipnoises}
\end{align}
i.e., ``independent noises'', simple relay policies are optimal in~\eqref{eq:capinidinputs}, i.e., the optimal probability mass function $\mathbb{P}_{S_{[1:N]}}$ has at most $N+1$ non-zero entries / active states.
\label{thm:main}
\end{thm}
\label{eq:allassumptions}
\end{subequations}

We first 
summarize some 
properties of submodular functions and LPs in Section~\ref{subsec:subLP},
we then prove Theorem~\ref{thm:main} in Section~\ref{subsec:Thproof},
we discuss the computational complexity of finding optimal simple schedules in Section~\ref{subsec:complexity}
and conclude with an example of a network with $N=2$ relays in order to illustrate some of the steps in the proof in Section~\ref{subsec:example}.

\subsection{Submodular Functions, LPs and Saddle-point Property} 
\label{subsec:subLP}

The following are standard results in submodular function optimization~\cite{BachSubm} and LPs~\cite{ChvatalLP}.

\begin{defin}[Submodular function, Lov\'{a}sz extension and greedy solution for submodular polyhedra]
\label{def:subm}
A set-function $f: 2^{N} \rightarrow \mathbb{R}$ is submodular if and only if, for all subsets $\mathcal{A}_1,\mathcal{A}_2 \subseteq [1:N]$, we have $f\left( \mathcal{A}_1\right)+f\left( \mathcal{A}_2\right) \geq f \left( \mathcal{A}_1 \cup \mathcal{A}_2 \right) + f \left( \mathcal{A}_1 \cap \mathcal{A}_2 \right)$
\footnote{A set-function $f$ is supermodular if and only if $-f$ is submodular, and
it is modular if it is both submodular and supermodular.}. Note that submodular functions are closed under non-negative linear combinations.

For a submodular function  $f$ such that $f(\emptyset) = 0$, the Lov\'{a}sz extension is a function defined as 
\begin{align}
\widehat{f} \left( \mathbf{w}\right) : = \max_{\mathbf{x} \in P(f)} \mathbf{w}^T \mathbf{x}, \quad  \forall\mathbf{w}  \in  \mathbb{R}^N,
\label{eq:lovext}
\end{align}
where $P(f)$ is the submodular polyhedron defined as 
\begin{align*}
P(f):= \left\{ \mathbf{x} \in  \mathbb{R}^N : \sum_{i \in \mathcal{A}} x_i  \leq  f(\mathcal{A}), \ \forall \mathcal{A} \subseteq [1:N] \right\}.
\end{align*}
The optimal $\mathbf{x}$ in~\eqref{eq:lovext} can be found by the greedy algorithm for submodular polyhedra and has components 
\begin{align*}
x_{\pi_i}= f\left( \{\pi_1, \ldots , \pi_i\}\right) -f\left( \{\pi_1, \ldots , \pi_{i-1}\}\right), \forall i \in [1:N],
\end{align*}
where $\pi$ is a permutation of $[1:N]$ such that the weights $\mathbf{w}$ are ordered as $w_{\pi_1} \geq w_{\pi_2} \geq \ldots \geq w_{\pi_N}$.  Note that the Lov\'{a}sz extension is a piecewise linear convex function.
\end{defin}

\begin{prop}[Minimum of submodular functions]
\label{prop:minsub}
Let $f$ be a submodular function such that $f(\emptyset) = 0$ and $\widehat{f}$ its Lov\'{a}sz extension.
The minimum of the submodular function satisfies
\begin{align*}
\min_{\mathcal{A} \subseteq [1:N]} f \left( \mathcal{A} \right) = \min_{\mathbf{w} \in [ 0:1]^N } \widehat{f} \left( \mathbf{w} \right)= \min_{\mathbf{w} \in [ 0,1]^N } \widehat{f} \left( \mathbf{w} \right),
\end{align*}
i.e., $\widehat{f} \left( \mathbf{w} \right)$ attains its minimum at a vertex of $[0,1]^N$.
\end{prop}

\begin{defin}[Basic feasible solution]
\label{def:BFS}
Consider the LP
\begin{align*}
\begin{array}{ll}
{\rm{maximize}} & \mathbf{c}^T \mathbf{x} 
\\ {\rm{subject \ to}} & \mathbf{A} \mathbf{x} \leq \mathbf{b} \quad \mathbf{x} \geq 0,
\end{array}
\end{align*}
where $\mathbf{x}\in  \mathbb{R}^n$ is the vector of unknowns, 
$\mathbf{b}\in  \mathbb{R}^m$ and $\mathbf{c}\in  \mathbb{R}^n$ are vectors of known coefficients,
and $\mathbf{A}\in  \mathbb{R}^{m \times n}$ is a known matrix of coefficients.
If $m<n$, a solution for the LP with at most $m$ non-zero values is called a basic feasible solution.
\end{defin}

\begin{prop}[Optimality of basic feasible solutions]
\label{th:optsolextr}
If a LP is feasible, then an optimal solution is at a vertex of the (non-empty and convex) feasible set $S= \left \{ \mathbf{x} \in \mathbb{R}^{n}: \mathbf{A}\mathbf{x} \leq \mathbf{b}, \mathbf{x} \geq 0\right \}$.
Moreover, if there is an optimal solution, then an optimal basic feasible solution exists as well.
\end{prop}

\begin{prop}[Saddle-point property]
\label{prop:minmaxeq}
Let $\phi(x,y)$ be a function of two vector variables $x \in \mathcal{X}$ and $y \in \mathcal{Y}$. By  the minimax inequality we have
\begin{align*}
d^{*}:= \max_{y \in \mathcal{Y}}\min_{x \in \mathcal{X}} \phi \left( x,y\right) \leq \min_{x \in \mathcal{X}} \max_{y \in \mathcal{Y}} \phi \left( x,y\right):=p^*.
\end{align*}
and equality holds, i.e., $p^*=d^*$, if
(i) $\mathcal{X}$ and $\mathcal{Y}$ are both convex and one of them is compact;
(ii) $\phi \left( x,y\right)$ is convex in $x$ and concave in $y$;
(iii) $\phi \left( x,y\right)$ is continuous.
\end{prop}

\subsection{Proof of Theorem \ref{thm:main}}
\label{subsec:Thproof}
The objective is to show that simple relay policies are optimal in~\eqref{eq:capinidinputs}.
The proof consists of the following steps:
\begin{enumerate}
\item
We first show that the function $I_{\mathcal{A}}^{(\text{fix})}$ defined in~\eqref{eq:IAfixed} is submodular under the assumptions in \eqref{eq:allassumptions}. 
\item
By using Proposition~\ref{prop:minsub}, we show that the problem in~\eqref{eq:capinidinputs} can be recast into an equivalent max-min problem.
\item
With Proposition~\ref{prop:minmaxeq} we show that the  max-min~problem is equivalent to solve a min-max~problem.
The min-max~problem is then shown to be equivalent to solve $N!$ max-min problems, for each of which we obtain an optimal basic feasible solution by Proposition~\ref{th:optsolextr} with the claimed maximum number of non-zero entries.
\end{enumerate}

\paragraph*{STEP 1}
%
We show that $I_{\mathcal{A}}^{(\text{fix})}$ in~\eqref{eq:IAfixed} is submodular.
The result in~\cite[ Theorem 1]{ParvaEtk} showed that $f_s(\mathcal{A})$ in~\eqref{eq:fs} is submodular for each relay state $s\in  [0:1]^N$ under the assumption of independent inputs and independent noises (the same work provided an example of a diamond network with correlated  inputs, and showed that in this case the cut-set bound is neither submodular nor supermodular). Since submodular functions are closed under non-negative linear combinations (see Definition~\ref{def:subm}), this implies that $I_{\mathcal{A}}^{(\text{fix})}  = \sum_{s  \in  [0:1]^N} \lambda_s \ f_s(\mathcal{A})$ is submodular under the assumptions of Theorem~\ref{thm:main}.

\begin{figure*}
\begin{align}
\min_{\mathcal{A} \subseteq [1:N]} I_{\mathcal{A}}^{(\text{fix})}
&= I_{\emptyset}^{(\text{fix})} + \min_{\mathcal{A} \subseteq [1:N]} g\left( \mathcal{A} \right) \nonumber
\\&
= I_{\emptyset}^{(\text{fix})} + \min_{\mathbf{w}\in [0,1]^N} 
\begin{bmatrix} w_{\pi_1} & w_{\pi_2} & \ldots & w_{\pi_N} \\ \end{bmatrix}
\begin{bmatrix} 
g\left( \{\pi_1\} \right) - g\left( \emptyset \right) \\ 
\vdots \\ 
g\left( \{\pi_1, \ldots , \pi_N\} \right) -g\left( \{\pi_1, \ldots , \pi_{N-1}\} \right) \\ \end{bmatrix} \nonumber
\\&= I_{\emptyset}^{(\text{fix})} +  \min_{\mathbf{w}\in [0,1]^N}
\begin{bmatrix} w_{\pi_1} & w_{\pi_2} & \ldots & w_{\pi_N} \\ \end{bmatrix}
\begin{bmatrix} 
I_{\{\pi_1\}}^{(\text{fix})}- I_{\emptyset}^{(\text{fix})} \\ 
\vdots \\ 
I_{\{\pi_1, \ldots , \pi_N\}}^{(\text{fix})} -I_{\{\pi_1, \ldots , \pi_{N-1}\}}^{(\text{fix})} \\ \end{bmatrix} \nonumber
\\&=   \min_{\mathbf{w}\in [0,1]^N}
\begin{bmatrix} 1 & w_{\pi_1} & w_{\pi_2} & \ldots & w_{\pi_N} \\ \end{bmatrix}
\begin{bmatrix} 
I_{\emptyset}^{(\text{fix})} \\
I_{\{\pi_1\}}^{(\text{fix})}  - I_{\emptyset}^{(\text{fix})} \\ 
\vdots \\ 
I_{\{\pi_1, \ldots , \pi_N\}}^{(\text{fix})} -I_{\{\pi_1, \ldots , \pi_{N-1}\}}^{(\text{fix})} \\ \end{bmatrix} 
:= \min_{\mathbf{w}\in [0,1]^N} \left\{ [ 1, \mathbf{w}^T ] \ \mathbf{H}_{\pi,f} \right\};
\label{eq:optle}
\\
\mathbf{H}_{\pi,f}
&:=
\mathbf{P}_{\pi} \
\underbrace{
\begin{bmatrix}
 1 & 0 & 0 & \ldots & 0\\
-1 & 1 & 0 & \ldots & 0\\
 0 &-1 & 1 & \ldots & 0\\
\vdots \\
 0 & 0& \ldots &-1 & 1\\
\end{bmatrix} 
}_{(N+1) \times (N+1)} \mathbf{F}_{\pi}
\label{eq:defHpif}
\ \text{where} \
\mathbf{F}_{\pi}:=
\begin{bmatrix}
f_0(\emptyset)         & \ldots & f_{2^N-1}(\emptyset)         \\
f_0(\{\pi_1\})             & \ldots & f_{2^N-1}(\{\pi_1\})             \\
f_0(\{\pi_1,\pi_2\})       & \ldots & f_{2^N-1}(\{\pi_1,\pi_2\})       \\
\ldots                       \\
f_0(\{\pi_1, \ldots, \pi_N\}) & \ldots & f_{2^N-1}(\{\pi_1, \ldots, \pi_N\}) \\
\end{bmatrix};
\end{align}
\begin{align}
\label{eq:leN2}
{ g \left( \mathcal{A} \right) = I_{\mathcal{A}}^{(\text{fix})} - I_{\emptyset}^{(\text{fix})},  \mathcal{A}\subseteq[1:2] : \quad }
\widehat{g}(w_1,w_2) = \left \{ \begin{array}{cc}  
  w_1 g \left( \left \{1 \right \}\right) 
+ w_2 \left[  g \left( \left \{1,2 \right \}\right) - g \left( \left \{1 \right \}\right)\right] & \text{if} \ w_1 \geq w_2 
\\ 
 w_2 g \left( \left \{2 \right \}\right )
+ w_1 \left[  g \left( \left \{1,2 \right \}\right) - g \left( \left \{2\right \}\right)\right] & \text{if} \ w_2 \geq w_1 \end{array} \right.;
\end{align}
\begin{align}
P_1: \max_{{{\bf \lambda}_{\rm{vect}}}} \min_{0\leq w_2\leq w_1 \leq 1} 
\underbrace{
\begin{bmatrix} 1 & w_1 & w_2 \end{bmatrix} \!\!
\begin{bmatrix}
 1 & 0 & 0 \\
 0 & 1 & 0\\
 0 & 0 & 1\\
\end{bmatrix} \!\!
\begin{bmatrix}
 1 & 0 & 0 \\
-1 & 1 & 0\\
 0 &-1 & 1\\
\end{bmatrix} \!\!
}_{=\begin{bmatrix} 1-w_1 & w_1-w_2 & w_2 \end{bmatrix}}
\underbrace{
\begin{bmatrix}
f_0(\emptyset) & f_1(\emptyset) & f_{2}(\emptyset) & f_{3}(\emptyset) \\
f_0(\{1\})         & f_1(\{1\})         & f_{2}(\{1\})         & f_{3}(\{1\})         \\
f_0(\{1,2\})       & f_1(\{1,2\})       & f_{2}(\{1,2\})       & f_{3}(\{1,2\})       \\
\end{bmatrix}}_{\mathbf{F}_{\pi^I}}\!\!
\begin{bmatrix} \lambda_0 \\ \lambda_1 \\ \lambda_2 \\ \lambda_3  \\ \end{bmatrix};
\label{eq:capacequivalent N=2 w1>w2}
\end{align}
\begin{align}
\begin{array}{lll}
P_2: &{\rm{maximize}} & \tau \\
     &{\rm{subject \ to}} 
     & \tau  \leq f_0(\emptyset)\lambda_0 + f_1(\emptyset)\lambda_1 + f_{2}(\emptyset)\lambda_2 + f_{3}(\emptyset)\lambda_3, \\
    && \tau  \leq f_0(\{1\})\lambda_0 + f_1(\{1\})\lambda_1 + f_{2}(\{1\})\lambda_2 + f_{3}(\{1\})\lambda_3, \\
    && \tau  \leq f_0(\{1,2\})\lambda_0 + f_1(\{1,2\})\lambda_1 + f_{2}(\{1,2\})\lambda_2 + f_{3}(\{1,2\})\lambda_3, \\
    && \lambda_0 + \lambda_1 + \lambda_2 + \lambda_3 = 1, \quad \lambda_i\geq 0 \ i\in[0:3]
\end{array};
\label{eq:P2equiv N=2 w1>w2}
\end{align}
\end{figure*}

\paragraph*{STEP 2}
Given that $I_{\mathcal{A}}^{(\text{fix})}$ in \eqref{eq:IAfixed} is submodular, we would like to use Proposition~\ref{prop:minsub} to `replace' the minimization over the subsets of $[1:N]$ in~\eqref{eq:capinidinputs} with a minimization over the cube $[0:1]^N$. Since $I_{\emptyset}^{(\text{fix})} = I \left( X_{[1:N+1]}; Y_{N+1}| S_{[1:N]} \right)\geq 0$ in general, we define a new submodular function $g\left( \mathcal{A} \right) := I_{\mathcal{A}}^{(\text{fix})} - I_{\emptyset}^{(\text{fix})}$ and proceed as in~\eqref{eq:optle} at the top of the next page to show that the problem in~\eqref{eq:capinidinputs} is equivalent to
\begin{align}
\mathsf{C}^\prime = 
\max_{{\bf \lambda}_{\rm{vect}}} \min_{\mathbf{w}\in [0,1]^N} \Big\{ [ 1, \mathbf{w}^T ] \ \mathbf{H}_{\pi,f} {\bf \lambda}_{\rm{vect}} \Big\},
\label{eq:capacequivalent}
\end{align}
where ${\bf \lambda}_{\rm{vect}}$ is the probability mass function of $S_{[1:N]}$
(in particular,  ${\bf \lambda}_{\rm{vect}}:=[\lambda_s] \in \mathbb{R}_+^{2^N\times 1}$ where $\lambda_{s}:= \mathbb{P}[S_{[1: N]}=s] \in[0,1]$, for $s\in [0:1]^N$ such that ${\sum_{s\in [0:1]^N} }\lambda_{s}   =   1$),
$\mathbf{H}_{\pi,f}\in \mathbb{R}^{(N+1) \times 2^N}$ and $\mathbf{F}_{\pi} \in \mathbb{R}^{(N+1) \times 2^N}$ are defined in~\eqref{eq:defHpif} at the top of the next page,
$\mathbf{P}_{\pi}  \in  \mathbb{R}^{(N + 1) \times (N + 1)}$ is the permutation matrix that maps
$[ 1 , w_{1} , \ldots , w_{N}]$ into $[1 , w_{\pi_1} , \ldots , w_{\pi_N}]$, and
$f_s \left( \mathcal{A}\right)$ was defined in~\eqref{eq:fs}.
We thus express our original optimization problem as the max-min~problem in~\eqref{eq:capacequivalent}.

\paragraph*{STEP 3}
In order to solve~\eqref{eq:capacequivalent} we would like to reverse the order of $\min$ and $\max$.
We note that the function $\phi \left( {\bf \lambda}_{\rm{vect}}, \mathbf{w}\right) := [ 1, \mathbf{w}^T ] \ \mathbf{H}_{\pi,f} {\bf \lambda}_{\rm{vect}}$ satisfies the properties in Proposition~\ref{prop:minmaxeq} (it is continuous, convex in $\mathbf{w}$ by the convexity of the Lov\'{a}sz extension and linear, thus concave, in ${\bf \lambda}_{\rm{vect}}$; moreover the optimization domain in both variables is compact).
Thus, we now focus on the problem
\begin{align}
\mathsf{C}^\prime=\min_{\mathbf{w}\in [0,1]^N} \max_{{\bf \lambda}_{\rm{vect}}}
\Big\{ [ 1, \mathbf{w}^T ] \ \mathbf{H}_{\pi,f} {\bf \lambda}_{\rm{vect}} \Big\},
\end{align}
which can be equivalently rewritten as
\begin{align}
\mathsf{C}^\prime
&=\min_{\pi \in \mathcal{P}_N} \min_{\mathbf{w}_{\pi}\in [0:1]^N} \max_{{\bf \lambda}_{\rm{vect}}}
\Big\{ [ 1, \mathbf{w}_{\pi}^T ] \ \mathbf{H}_{\pi,f} {\bf \lambda}_{\rm{vect}} \Big\}
\label{eq:P1}
\\&=\min_{\pi \in \mathcal{P}_N}  \max_{{\bf \lambda}_{\rm{vect}}}\min_{\mathbf{w}_{\pi}\in [0:1]^N}
\Big\{ [ 1, \mathbf{w}_{\pi}^T ] \ \mathbf{H}_{\pi,f} {\bf \lambda}_{\rm{vect}} \Big\},
\label{eq:P2}
\end{align}
where $\mathcal{P}_N$ is the set of all the $N!$ permutations of $[1:N]$.
In~\eqref{eq:P1}, for each permutation $\pi \in \mathcal{P}_N$, 
we first find the optimal ${\bf \lambda}_{\rm{vect}}$, 
and then find the optimal $\mathbf{w}_{\pi} : w_{\pi_1} \geq w_{\pi_2} \geq \ldots w_{\pi_N}$.
This is equivalent to~\eqref{eq:P2}, where again by Proposition~\ref{prop:minmaxeq}, 
for each permutation $\pi \in \mathcal{P}_N$, 
we first find the optimal  $\mathbf{w}_{\pi} : w_{\pi_1} \geq w_{\pi_2} \geq \ldots w_{\pi_N}$,
and then find the optimal ${\bf \lambda}_{\rm{vect}}$. 

Let's now consider the inner optimization in~\eqref{eq:P2}, that is, the problem
\begin{align}
P_1: 
\max_{{\bf \lambda}_{\rm{vect}}} \min_{\mathbf{w}_{\pi}\in [0:1]^N}  \Big\{ [ 1, \mathbf{w}_{\pi}^T ] \ \mathbf{H}_{\pi,f} {\bf \lambda}_{\rm{vect}} \Big\}.
\label{eq:newOptPrDual}
\end{align}

From Proposition \ref{prop:minsub} we know that, for a given $\pi \in \mathcal{P}_N$, the optimal $\mathbf{w}_{\pi}$ is a vertex of the cube $[0:1]^N$. For a given ${\pi} \in \mathcal{P}_N$, there are $N+1$ vertices whose coordinates are ordered according to $\pi$. 
%
In~\eqref{eq:newOptPrDual}, for each of the $N+1$ feasible vertices of $\mathbf{w}_{\pi}$, it is easy to see that the product $[ 1, \mathbf{w}_{\pi}^T ] \ \mathbf{H}_{\pi,f}$ is equal to a row of the matrix $\mathbf{F}_{\pi}$. By considering all possible $N+1$ feasible vertices compatible with $\pi$ we obtain all the $N+1$ rows of the matrix $\mathbf{F}_{\pi}$.
Hence, 
$P_1$ is equivalent to
\begin{align}
\begin{array}{lll}
P_2:&{\rm{maximize}} & \tau 
\\ &{\rm{subject \ to}} &  \mathbf{1}_{(N+1)} \tau  \leq \mathbf{F}_{\pi} {\bf \lambda}_{\rm{vect}}
\\ &{\rm{and}} &  \mathbf{1}_{2^N}^T  {\bf \lambda}_{\rm{vect} } = 1, \ {\bf \lambda}_{\rm{vect} }\geq 0.
\end{array}
\label{eq:polmnuytf}
\end{align}
The LP $P_2$ has 
$n=2^N+1$ optimization variables ($2^N$ values for ${\bf \lambda}_{\rm{vect}}$ and $1$ value for $\tau$),
$m=N+2$ constraints, and 
is feasible (consider for example 
the uniform distribution of ${\bf \lambda}_{\rm{vect}}$ and  $\tau=0$).
Therefore, by Proposition~\ref{th:optsolextr}, $P_2$ has an optimal basic feasible solution with at most $m=N+2$ non-zero values. Since $\tau >0$ (otherwise the channel capacity would be zero), it means that ${\bf \lambda}_{\rm{vect}}$ has at most $N+1$ non-zero entries.

Since for each $\pi \in \mathcal{P}_N$ the optimal ${\bf \lambda}_{\rm{vect}}$ in~\eqref{eq:P2} has at most $N+1$ non-zero values, then also for the optimal permutation the corresponding optimal ${\bf \lambda}_{\rm{vect}}$ has at most $N+1$ non-zero values. This shows that the optimal schedule in the original problem in~\eqref{eq:capinidinputs} is simple. 
This concludes the proof of Theorem \ref{thm:main}.

%

\subsection{On the complexity of finding the optimal simple schedule}
\label{subsec:complexity}
Our proof method seems to suggest that finding the optimal schedule requires the solution of $N!$ LPs.
Since $\log(N!) = O(N\log(N/e))$, the computational complexity of this approach would be prohibitive for large $N$.
By using an iterative method that alternates between the submodular function minimization over $\mathbf{w}$ ({by using the Schrijver's algorithm, this is solvable in strongly polynomial time in $N$) and the LP maximization over ${\bf \lambda}_{\rm{vect}}$ (by the ellipsoid method, the worst-case dual LP is solvable in polynomial time in $N$)
we have a polynomial time algorithm that converges to the optimal solution by the saddle-point property, which holds with equality for our problem.



\subsection{Examples} 
\label{subsec:example}
For $N=2$, $P_1$ in~\eqref{eq:newOptPrDual} requires an optimization over $\mathbf{w}=[w_1,w_2]\in [0,1]^2$. 
From Proposition \ref{prop:minsub}, the optimal $\mathbf{w}$ 
is one of the vertices $(0,0),(0,1),(1,0),(1,1)$.
We must consider $|\mathcal{P}_2|= 2!=2$ possible permutations: 
$\pi^I$ 
for which $w_1\geq w_2$, and 
$\pi^{II}$ 
for which $w_2\geq w_1$.
For $N=2$ the Lov\'{a}sz extension of a submodular function $g$ 
is given in~\eqref{eq:leN2} at the top of the previous page (see also $\rm{eq.}\eqref{eq:optle}$), which results in the problem $P_1$ in~\eqref{eq:capacequivalent N=2 w1>w2} at the top of the previous page when considering $\mathbf{w}_{\pi^I}$ (a similar reasoning holds for $\mathbf{w}_{\pi^{II}}$ but it is not reported here for sake of space).
%
The vertices compatible with $\pi^I$ are 
$[w_1,w_2]\in\{(0,0),(1,0),(1,1) \}$, 
which result in $[1-w_1, w_1-w_2, w_2]\in\{(0,0,0),(0,1,0),(0,0,1) \}$.
This implies that $P_2$ in~\eqref{eq:P2equiv N=2 w1>w2} at the top of the previous page is the minimum of three functions, each given by one of the rows of $\mathbf{F}_{\pi_I}$ 
multiplied by ${{\bf \lambda}_{\rm{vect}}}=[\lambda_0,\lambda_1,\lambda_2,\lambda_3]$. 
%
%
Therefore, $P_2$ has $4$ constraints ($3$ from the rows of $\mathbf{F}_{\pi_I}$ and $1$ from ${{\bf \lambda}_{\rm{vect}}}$) and $5$ unknowns ($1$ value for $\tau$ and $4$ entries of ${{\bf \lambda}_{\rm{vect}}}$). Thus, by Proposition~\ref{th:optsolextr}, $P_2$ has an optimal basic feasible solution with at most $4$ non-zero values, of which $1$ is $\tau$ and thus the other (at most) $3$ belong to ${{\bf \lambda}_{\rm{vect}}}$. 
By the work of~\cite{Bagheri2009} and our generalization in~\cite{ourITjournal}, we know that either $\lambda_0$ or $\lambda_3$ is zero, thus giving the desired optimal simple schedule.

As mentioned earlier, the result of this paper proves our original conjecture in~\cite{ourITW2014} for  Gaussian SISO networks for any number $N$ of relays.
Our framework also immediately extends to Gaussian networks with MIMO relays and independent noises since also in this setting independent inputs at all nodes are optimal in the cut-set upper bound to within a constant gap for all choices of the channel matrices.

\section{Conclusions}
\label{sec:Concl}
In this work we studied networks with $N$ half-duplex relays. For such networks, the capacity achieving scheme must be optimized over the $2^N$ possible listen-transmit relay configurations. This paper formally proved that, if noises are independent and independent inputs are approximately optimal in the cut-set bound, then
the approximately optimal schedule only uses $N+1$ relay configurations.

{\bf Acknowledgments:}
This work received fundings from Eurecom's industrial partners,
the EU Celtic+ Framework Program Project SHARING, and from a 2014 Qualcomm Innovation Fellowship.
The work of D.~Tuninetti was partially funded by NSF under award number 1218635; the contents of this article are solely the responsibility of the author and do not necessarily represent the official views of the NSF.
D.~Tuninetti would like to acknowledge insightful discussions with Dr. Salim~El~Rouayheb on sumbodular functions.

\bibliographystyle{IEEEtran}
\bibliography{ITWBib}

\end{document}